International Conference on Mobile Money Uptake, by Ghana Technology University College (GTUC) in partnership with the Institute for Money, Technology and Financial Inclusion (IMTFI) USA, at Accra, Ghana, 2013

# The Role of Rural Banks in Providing Mobile Money Services to Rural Poor Communities: An effective integration approach of Rural Banks and existing mobile communications infrastructure.

Quist-Aphetsi Kester, MIEEE

Lecturer, Faculty of Informatics Ghana Technology University College, Accra, Ghana

Email: kquist-aphetsi@gtuc.edu.gh

*Abstract* - *The rapid spread of mobile phones means that the number of mobile users may already exceed the number of banked people in many low income countries. Mobile phones can also offer a communications channel for initiating and executing on-line financial transactions. This channel may not only reduce the cost of financial transactions for provider and customer, but also allow new entrants to the financial sector, and new relationships to be formed for distributing services. These changes hold the prospect of accelerating access to financial services on the back of the mobile infrastructure.*

*Mobile telephony offers tremendous promise to facilitate the flow of money among rural and poor families at much lower transaction costs, bringing the bank to those currently unbanked. Realizing this promise will require close collaboration among all stakeholders.*

*But most rural banks do not have mobile banking services for their customers. This made it difficult for the full potential and benefits of mobile money financial services to be realized. Most telecommunication service providers run mobile money service solely for their subscribers without an integrated approach of incorporating and integrating rural banking systems into their existing services this makes it difficult for a full fledge exploitation of the mobile financial market.*

*This paper looks at the existing mobile money services and takes critical look at the positive advantages of effective integration approach of Rural Banks and existing mobile communications infrastructure as well as proposing a model for such integration.*

International Conference on Mobile Money Uptake, by Ghana Technology University College (GTUC) in partnership with the Institute for Money, Technology and Financial Inclusion (IMTFI) USA, at Accra, Ghana, 2013

## I. Introduction

The rise of mobile network operators globally opened the door for mobile applications that render services for people across geographical areas. They provided services such as text messaging, MMS, email, Internet access, short-range wireless communications (infrared, Bluetooth), business applications, gaming, mobile money transfer, photography and others.

Mobile payment, also known as mobile money, mobile money transfer, and mobile wallet generally refer to payment services operated under financial regulation and performed from or via a mobile device. Instead of using cash, check, or credit cards, a consumer can use a mobile phone to pay for a wide range of services and digital or hard goods.

According to Cameron Peake (2012), there are more than 130 live mobile money deployments tracked globally by the GSMA, the mobile telecom industry body, and another 87 in development. 2 Of bank-led initiatives, there are 236 agent banking deployments in Brazil, Peru, Colombia and Mexico alone, with a total of more than 43,000 combined agents. As the market is further defined and developed, payment actors such as Visa, MasterCard and Western Union are positioning in this space as well.

With reference to a publication by Daily Guide (2011), the International Telecommunications Union estimates mobile subscriptions across Africa have more than tripled to 333 million since 2005 and in South Africa, the DRC, Zambia and Kenya, mobile phone banking has offered services to remote areas where conventional banks have been physically absent.

Mobile telephony offers tremendous opportunities for rural communities in developing countries and have promising goal to facilitate the flow of money among rural and poor families at much lower transaction costs, bringing the bank to those currently unbanked. Realizing this promise will require close collaboration among all stakeholders.

In the midst of these prospects that mobile money offers, most rural banks do not have mobile banking services for their customers, currently. This made it difficult for the full potential and benefits of mobile money financial services to be realized in rural communities. One major challenge also is that most telecommunication service providers run mobile money service solely for their subscribers without an integrated approach of incorporating and integrating rural banking systems into their existing services, and this makes it difficult for a full fledge exploitation of the mobile financial market.

This paper looks at the existing mobile money services and takes critical look at the positive advantages of effective integration approach of Rural Banks and existing mobile communications infrastructure as well as proposing a platform model for such integration. The model proposed will use an integration approach that will interconnect Rural Banks and existing mobile communications infrastructure via a service bus based on the concept of Service oriented architecture. This will link mobile money service providers effectively as well as making possible future integration solution and interoperability of systems for effective service delivery.

The paper has the following structure: section II consist of related works, section III gives information on the methodology, section IV discusses the approach used for the integration, V talks about implementation as well as results and concluded the paper.

## II. Literature Review

According to Cameron Peake (2012), Infrastructure changes in rural areas are external factors that may force new models for the mobile services. Mobile phone and Internet penetration are typically lower, which may severely raise costs and/or inhibit the provider's ability to introduce mobile or Internet-based services. For bank-led models, it may be wise to team with an MNO for rural development or look at more human-centered solutions where transaction data may not be real time, but regularly synched as the network allows. One thing to keep in mind is that technology infrastructure is rapidly expanding, and though a system may not be in place today, it may be completely functional in six months.

With service oriented architecture and web services, mobile phones transactions will be platform independent and phones that do not have internet connectivity can still do banking with other systems that are internet or extranet dependent.

One other thing that Cameron Peake(2012) identified again was that, regulations also force new models to emerge for the mobile services. In countries where bank account opening is not allowed at the agent level and rural banking penetration is low, typical (that is, transactional) agent models will be of limited use, and the value proposition for rural clients will be reduced. Some banks may introduce roaming employees to register clients, thus avoiding the restrictions. It should also be kept in mind that agent regulation is rapidly expanding as governments become more familiar with nontraditional models.

With the proposed integrated systems, once a mobile customer has a registered mobile line and has been registered on the mobile service officially by a Mobile service provider, financial service transactions of banks will be made readily available based on service level contracts. Which means can receive and transfer money for any transactions that fit into the domain at the service contract level.

According to Jenkins, B. (2008), As Hans Wijayasuriya, CEO of Sri Lanka's Dialog Telekom, asserts, "for something to be ubiquitous, the foundations need to be very strong." 16 Without doubt, scalable, robust technology will be required. Fundamo's Hannes van Rensburg asks, "How do we build a system that would scale from hundreds of transactions a second to thousands a second? What does it do under adverse conditions and how quickly can it recover? What we can't afford in this infant industry is major catastrophes." 17 What other foundations do growing mobile money ecosystems need? Their leaders identify three that stand out: utility, capacity, and an enabling environment.

Jenkins, B. (2008) further explained that some degree of interoperability will be required if a critical mass of mobile money services is to develop. Mobile network operators and other companies are conscious of their core competencies and often, quite rightly, want to focus on their core businesses. As a result, some interoperability, in the sense of collaboration or partnership, will be needed to "expose [consumers] to a broad range of players so collectively they can get the full package of financial services." 28 As Alex Ibasco, Group Head of New Business Streams at Smart Communications, says, "Smart has no illusion it can build everything. Instead, we take a strategy of inclusion, inviting people to come in and create businesses out of areas where there are gaps. We think of ourselves as a horizontal infrastructure enabler and we look for people that think vertically. Creating businesses out of the gaps – that is the only way."

As mobile money services spread across geographical areas, there is a need for systems that can accommodate the rapid growth of transactions and the system should also have the capability of interoperability as well as accommodating future changes without a negative effect on the existing system. Hence this paper provides a solution to solving this problem.

### III. Methodology

As elaborated by Paul A. Strassmann(2007), SOA with web services changes the way businesses are undertaken and it is a technique of design that guides all aspects of creating and using business services throughout their lifecycle Thus, from their conception to their retirement. And also in defining and conditioning the IT infrastructure that allows different applications to exchange data and participate in business processes regardless of the operating systems or programming languages underlying those applications.

For mobile money to achieve its full potential, interoperability of systems services of different banks and telecommunication industries is the key. This research work focuses on the design of architectures to illustrate the model of integration between banking systems and communications infrastructure within the mobile money industry and then propose a model for its effective integration.

The paper seeks to achieve two main aims, the first aim is to integrate systems such that they can interoperate and provide services to each other, whereas the second aim is to provide an avenue for addition of new services in the future by interconnecting with other services without changing the entire systems. However in high hopes of achieving these aims, the existing IT infrastructure will not be replaced by a new one due to extensive investments made in the existing systems. It's more cost-effective to evolve and enhance the systems. The approach will also facilitate the composition of services across disparate pieces of software systems to streamline IT processes and eliminate barriers to IT environment improvements within and outside organizations operations.

International Conference on Mobile Money Uptake, by Ghana Technology University College (GTUC) in partnership with the Institute for Money, Technology and Financial Inclusion (IMTFI) USA, at Accra, Ghana, 2013

## IV. The approach

The approach used was to get all the independent systems residing within and outside each organization's system connected so that they can easily exchange data irrespective of the platform upon which they are running on and make the services available for usage by external systems during transactions. A multi-layer architecture approach was used to develop a model to make information processing possible within and outside the organizations and also making some services accessible. Back-end database servers were set up and Web services were modeled to make it possible for external and internal organizations to request and provide services through the service bus.

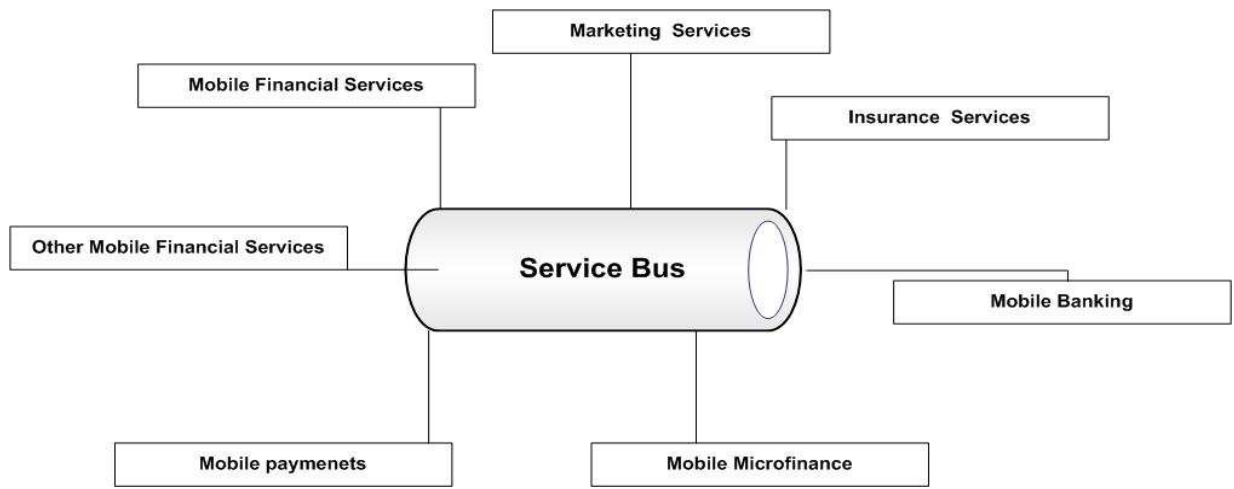

Figure 1: The Mobile Services Integration model

Mobile Services Integration model via a service bus: This model interconnects all services through a services bus by making it possible for service contracts to be made easily as well as providing a platform for easy integration, interoperability and addition of new layers of abstraction. Which means that phones with or without internet connectivity can perform the same transactions. Laptops, desktops, other mobile and non mobile devices with internet browsing capability can also have access to such services.

International Conference on Mobile Money Uptake, by Ghana Technology University College (GTUC) in partnership with the Institute for Money, Technology and Financial Inclusion (IMTFI) USA, at Accra, Ghana, 2013

Figure 2 showed Service Oriented Integrated architecture of Enterprise system resources. The Enterprise Resources and Operational Systems consist of existing applications, legacy and COTS systems, CRM and ERP applications, and older OO implementations. Integration Services provide access to the resources and systems of Enterprise Resources and Operational Systems and the components wrap integration services and provide a 'single point of contact' for integration services. Business Services represents a logical grouping of component, integration services and operations and also provides high level business functionality throughout the enterprise as well as provides a 'service interface' layer of abstraction to the functionality of components and Integration Services. Business Processes as a series of activities executed in an ordered sequence according to business events and a set of business rules called choreography or business process model.

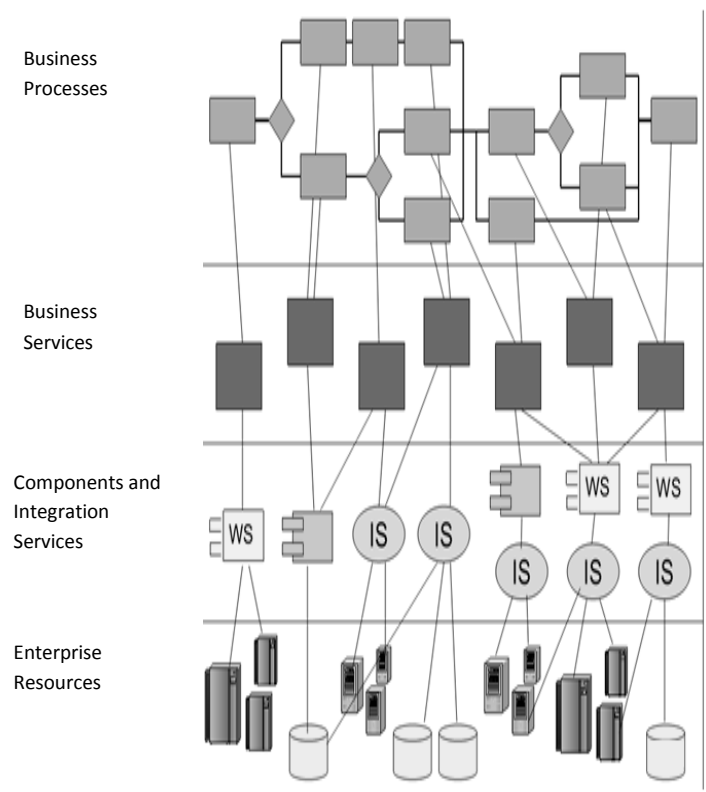
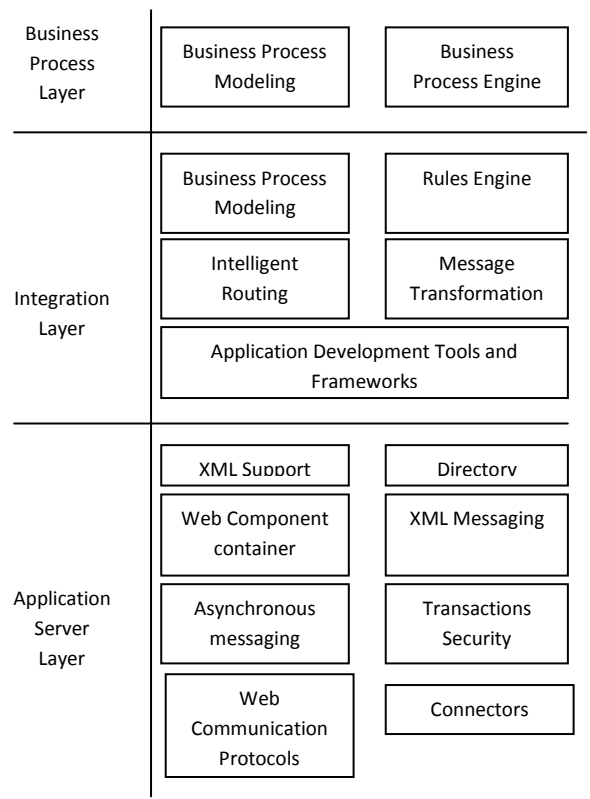

Figure 2: Service Oriented Integrated Architecture                Figure 3: Application Integration Layers

Figure 3 showed web services exposure through the interface functionality performed by the internal systems and this makes the services discoverable and accessible through the Web in a controlled manner as well as the service bus. Homogeneous components were modeled to reduce the difficulties of integration and standardized. From the architecture Service descriptions were made richer and more detailed, covering aspects beyond the service interface. The application integration environment encompasses three layers: a business process layer, an integration layer, and an application server layer. Each layer, in turn, holds technologies that serve as the application server integration building blocks. The application server layer enables an application integration project to link not only with existing enterprise systems but also with the Web. The application integration platform adds an integration layer on top of application server. This integration layer provides support for application development tools and frameworks. These development tools and integration frameworks are based on the application programming model, and they rely on metadata for generating and providing services. The integration layer also adds support for such functionality as a rules engine, intelligent message routing, and message transformation, all on top of the base functionality provided by the application server. The business process layer serves as the top-most layer for the platform and represents an enterprise's unique way of doing business. This business process layer exposes business process level abstraction by providing support for business process modeling and for the business process engine.

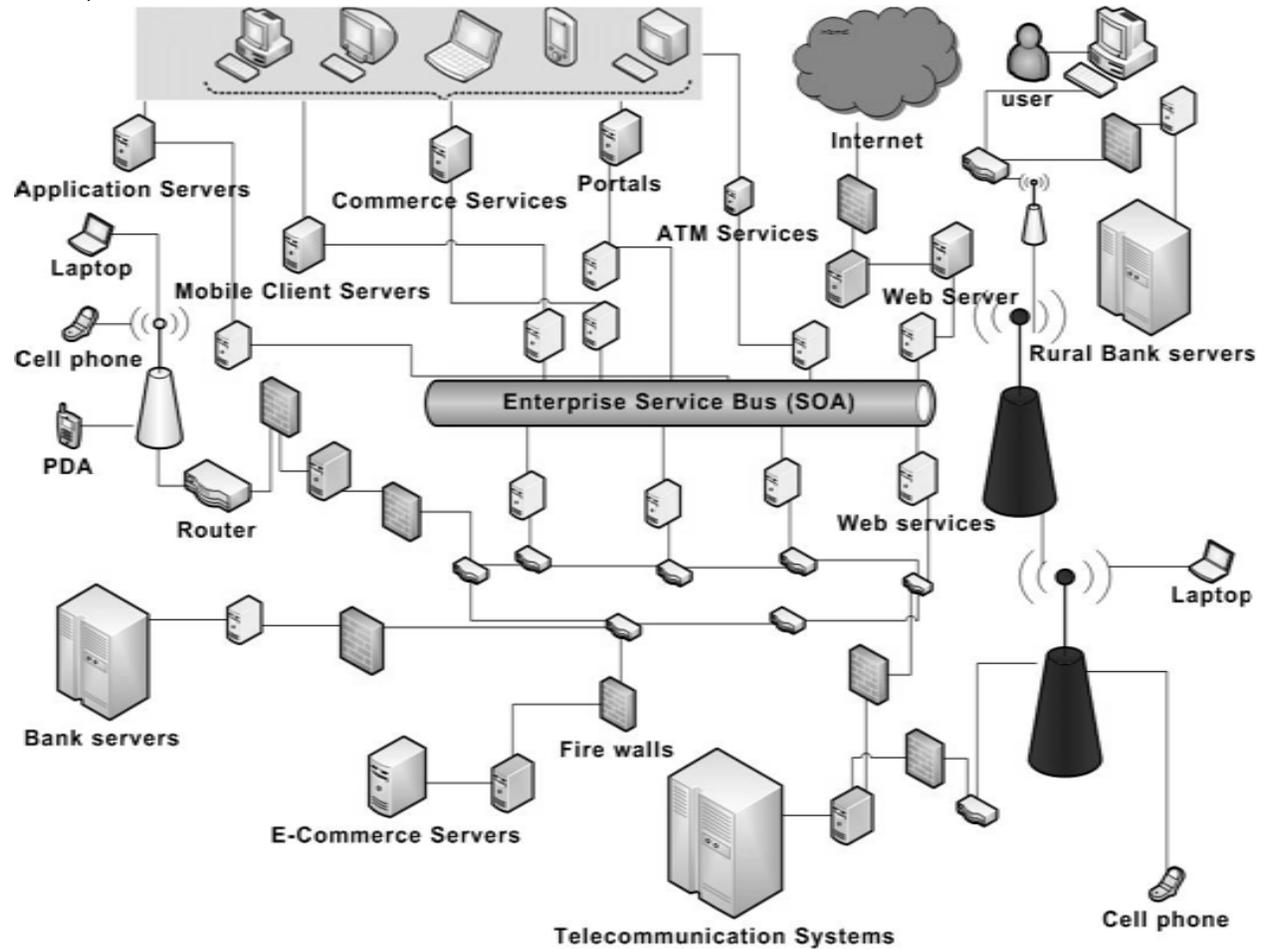

Figure 4: Integrated Systems architecture via a Service Bus.

The service bus implements, integrates, manages, shares and distribute the services. Through the service bus end users and heterogeneous devices, can interconnect, interoperate and provide services to each other irrespective of the end users' location, device platforms and programming languages. As the business logic layer is service-enabled, the presentation logic of a particular system can be separated from the business logic thereby making it easier to interconnect with various types of Graphical User Interfaces and devices. For the services to reflect and correlate with business processes, the business process engine exists at the business logic in the architecture to describe business processes, automate, modify and enforce business rules and drive an automatic flow of execution across the multiple services.

## V.     Conclusion

With the proposed model, architectures, SOA Concepts and Service oriented development approach, an effective integration approach of Rural Banks and existing mobile communications infrastructure can be achieved. This approach is less expensive to implement and has less work during the system integration stage.

Rural banks then can now implement this approach to exploit the full benefits of mobile banking. Money can be easily transferred from one account to the other from different rural banks account to the other. One can also now send money from one's mobile money account from a network say Vodafone to the other say MTN without going through any difficulty.

International and national transfer of money can easily take place for any form of mobile money transaction services. Payment can be done effectively using mobile money and withdrawals can be done from banks and any mobile money agent.

Future works will be based on how detailed implementation of the project can be done at the rural bank level for transactions to be done via an SMS gateway. This will help the banks to provide an avenue for the rural folks do business using money from their accounts at the rural banks at anytime.